\documentclass[twocolumn,showpacs]{revtex4}\newcommand{\JHEPonly}[1]{}\newcommand{\PRDonly}[1]{#1}
\usepackage{amsmath,amssymb,dcolumn}
\usepackage{graphicx}
\def\imo{i}

\JHEPonly{
\newenvironment{widetext}{}{}
\setlength{\textwidth}{16.3cm}

\title{\textbf{Long life of Gauss-Bonnet corrected black holes}}
\author{\textbf{R. A. Konoplya}\\
Department of Physics, Kyoto University, Kyoto 606-8501, Japan  \\
\& \\
Theoretical Astrophysics, Eberhard-Karls University of T\"{u}bingen, T\"{u}bingen 72076, Germany}
\author{A. Zhidenko\\
Instituto de F\'{\i}sica, Universidade de S\~{a}o Paulo \\ C.P. 66318, 05315-970, S\~{a}o Paulo-SP, Brazil}

\date{}

\abstract{ Dictated by the string theory and various higher dimensional scenarios, black holes in $D>4$-dimensional space-times must have higher curvature corrections. The first and dominant term is quadratic in curvature, and  called the Gauss-Bonnet (GB) term. We shall show that although the Gauss-Bonnet correction changes black hole's  geometry only softly, the emission of gravitons is suppressed by many orders even at quite small values of the GB coupling. The huge suppression of the graviton emission is due to the multiplication of the two effects: the quick cooling of the black hole when one turns on the GB coupling and the exponential decreasing of the grey-body factor of the tensor type of gravitons at small and moderate energies. At higher $D$ the tensor gravitons emission is dominant, so that the overall lifetime of black holes with Gauss-Bonnet corrections is many orders larger than was expected. This effect should be relevant for the future experiments at the Large Hadron Collider (LHC).}
}

\begin{document}

\PRDonly{
\title{Long life of Gauss-Bonnet corrected black holes}
\author{R. A. Konoplya}\email{konoplya_roma@yahoo.com}
\affiliation{Department of Physics, Kyoto University, Kyoto 606-8501, Japan\\
\& \\
\mbox{Theoretical Astrophysics, Eberhard-Karls University of T\"{u}bingen,}\\ T\"{u}bingen 72076, Germany}
\author{A. Zhidenko}\email{zhidenko@fma.if.usp.br}
\affiliation{Instituto de F\'{\i}sica, Universidade de S\~{a}o Paulo,\\
C.P. 66318, 05315-970, S\~{a}o Paulo-SP, Brazil}

\begin{abstract}
Dictated by the string theory and various higher dimensional scenarios, black holes in $D>4$-dimensional space-times must have higher curvature corrections.  The first and dominant term is quadratic in curvature, and  called the Gauss-Bonnet (GB) term. We shall show that although the Gauss-Bonnet correction changes black hole's  geometry only softly, the emission of gravitons is suppressed by many orders even at quite small values of the GB coupling. The huge suppression of the graviton emission is due to the multiplication of the two effects: the quick cooling of the black hole when one turns on the GB coupling and the exponential decreasing of the grey-body factor of the tensor type of gravitons at small and moderate energies. At higher $D$ the tensor gravitons emission is dominant, so that the overall lifetime of black holes with Gauss-Bonnet corrections is many orders larger than was expected. This effect should be relevant for the future experiments at the Large Hadron Collider (LHC).
\end{abstract}

\maketitle
}

\section{Introduction}

During the past decade high energy physics received a great impact from theories implying existence of extra dimensions in the world. These are the string theory \cite{strings} and higher dimensional brane-world scenarios \cite{BWS}. The low energy limit of the string theory can be described by the slope expansion in powers of the inverse string tension (or of the inverse square of the fundamental string scale $\ell_{s}^{-2}$) that produces higher curvature corrections to the Einstein action. The quadratic term in curvature (given by the so-called Gauss-Bonnet invariant) is the leading correction that can affect the graviton excitation spectrum near the flat space.

The extra dimensional scenarios also suggest that the fundamental gravity scale $M_{*}$ might be around the weak scale $\sim TeV$. Thus, at particle collisions with the cross section $\sim \pi r_{s}^{2}$, where $r_s$ is the Schwarzschild radius, and energies larger than $M_{*}$, the production of mini-black holes should start. These black holes are intrinsically higher dimensional and usually modeled by the Tangherlini metric, which is the solution of the D-dimensional Einstein equations. However, in order to have a mathematically noncontradictory gravity in higher dimensions, one has to take account of higher curvature corrections of the same form as those appearing in the slope expansion of the string theory. The spherically symmetric solution describing neutral static black holes in the D-dimensional Einstein gravity with the GB corrections was obtained in \cite{Deser}. This solution contains small corrections to the D-dimensional Schwarzschild-Tangherlini geometry and consequently properties of such Gauss-Bonnet corrected black holes were expected to differ only slightly from the Schwarzschild's ones. This happens, for instance, for the spectrum of proper oscillations of these black holes \cite{GBQNMs}.

Unlike astrophysical black holes, whose Hawking evaporation is negligibly small, mini-black holes are intensively evaporating what leads to the very short lifetime of these black holes, once they are created. The latter is due-to strong production of various particles from the vacuum around a black hole and emission of them through the mechanism of Hawking radiation \cite{Hawking:1974sw}. At large number of space-time dimensions $D$, the specific ``tensorial'' type of gravitons  (respectively the $D-2$ rotation group) dominates in the emission process \cite{Cardoso:2005vb}. Up to now, an impressively extensive literature is devoted to the calculations of Hawking evaporation of the Schwarzschild-Tangherlini and Myers-Perry black holes \cite{Jung-rot,Chen,charybdis2,IOP2,Nomura}, while evaporation of their higher curvature corrected generalizations was touched upon only in a couple of works \cite{Rizzo:2006uz,Grain:2005my}. In particular, T.~Rizzo estimated the energy-emission rate for the higher curvature corrected black holes, assuming that the grey-body factor equals unity \cite{Rizzo:2006uz}. This was expected to give the correct answer about the order of the intensity of the Hawking emission. Though, as we shall show in this paper, the contribution due to the grey-body factors can also considerably change the results. In \cite{Grain:2005my}, the scattering of Standard Model particles around Gauss-Bonnet black holes was considered, though the calculations were terminated at the grey-body factors and the numbers of particles per frequency. Thus, none of the above works calculated the energy-emission rate for Gauss-Bonnet black holes that is necessary for the estimation of the total emission of energy and thus of the black hole lifetime. Here we shall fill this gap and calculate the energy-emission rates for fields of various spin, including gravitons, and thus shall estimate the lifetime of Gauss-Bonnet black holes.

In this work, we shall show that due to a number of reasons, the emission of the tensor type of gravitons is greatly (in fact exponentially) suppressed when one turns on the GB coupling $\alpha'$. Thus, even at small values of the GB coupling constant $\alpha'$ \emph{the graviton emission is suppressed by many orders}. This means that small GB corrections lead to a much longer life of higher dimensional black holes than was expected \cite{Cardoso:2005vb,Kanti:2008eq,Kanti:2009sn}.
At first sight this enormous suppression would not seem trustworthy: why do slight corrections of geometry produce a very strong effect on the evaporation process? The reasons for this are ``multiplication'' of the two factors. First, the black hole gets much colder when one turns on the GB coupling and the emission rate is quadratic in temperature. Second, the emission is proportional to the grey-body factor which is exponentially suppressed for tensorial gravitons. This explanation certainly did not make us trust the result immediately. Therefore we reproduced our accurate numerical calculations by the semianalytical WKB estimations.

The paper is organized as follows: Sec \ref{sec:wave-like} briefly discuss the deduction of wave equations for perturbations of fields of various spin. Sec \ref{sec:methods} is devoted to numerical calculation of the coefficients of transmission, while Sec. \ref{sec:WKB} gives WKB values of the coefficients. In Sec. \ref{sec:results} the obtained scattering data are used for the calculations of the energy-emission rates. Using the WKB arguments Sec. \ref{sec:discussions} explains why the found enormous suppression of Hawking evaporation occurs. In Sec. \ref{sec:conclusions} we estimate the lifetime of Gauss-Bonnet corrected black holes and outline the future perspective for this direction.

We shall consider the canonical ensemble, which leads to the same results as the microcanonical one if the black hole mass $M$ is at least a few times larger than $M_*$ \cite{Rizzo:2006uz}.

\begin{figure*}
\includegraphics[width=.5\textwidth,clip]{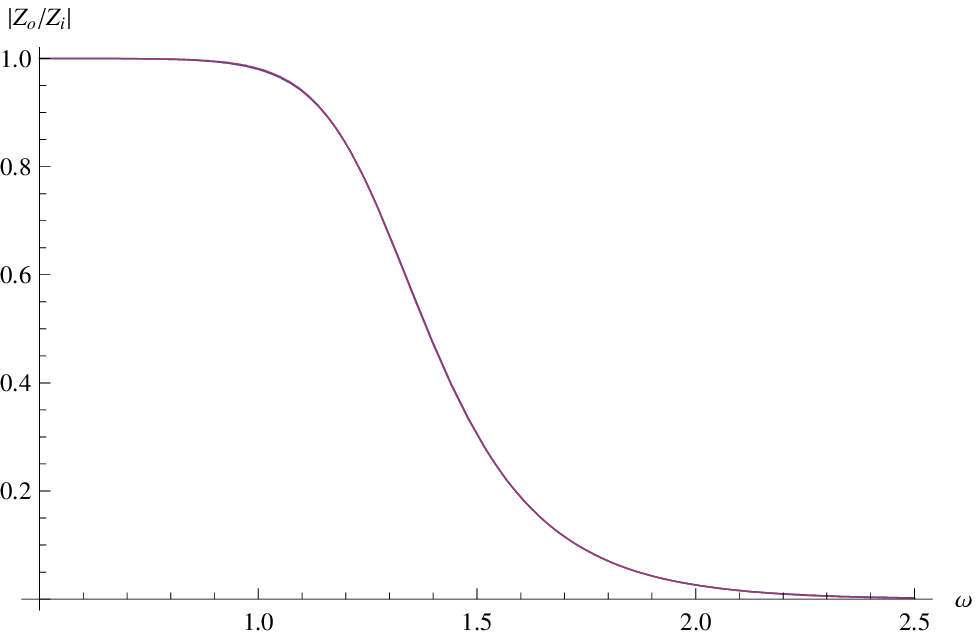}\includegraphics[width=.5\textwidth,clip]{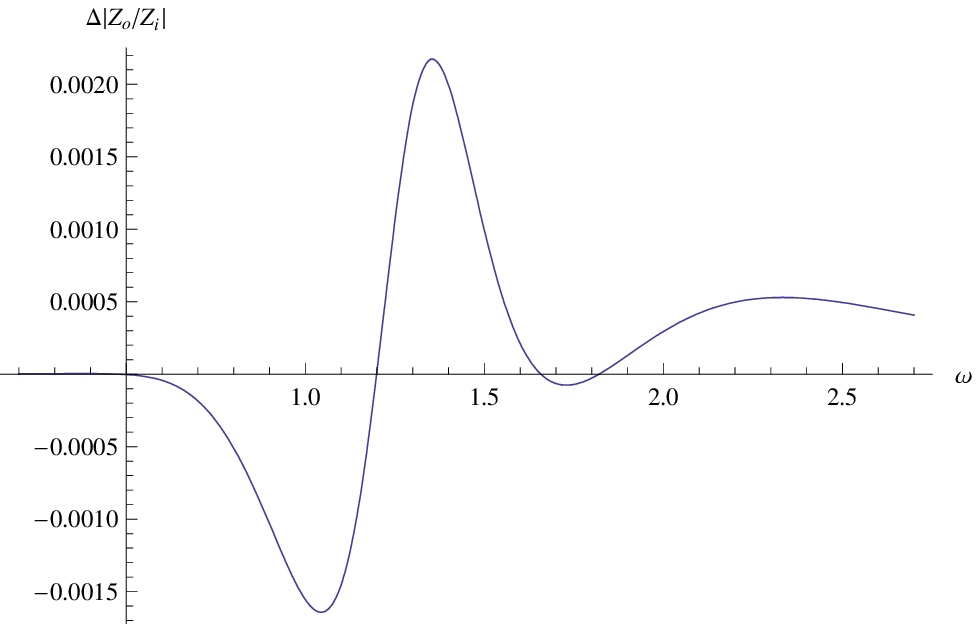}
\caption{Square root of the reflection coefficient $|Z_o/Z_i|$ for $D=6$, $\alpha=2$, $l=2$ tensor-type gravitational perturbations. The left panel shows the coefficient calculated by fitting the numerically solved equation (blue) and using the 6-th order WKB formula (red). In the right panel we plot the difference between the coefficients calculated using these two methods.}\label{fig-WKBcheck}
\end{figure*}

\begin{figure*}
\includegraphics[width=.5\textwidth,clip]{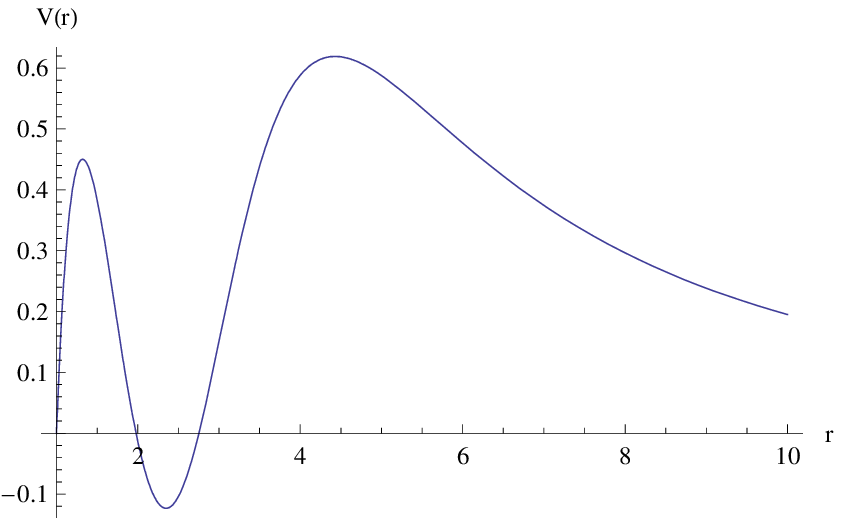}\includegraphics[width=.5\textwidth,clip]{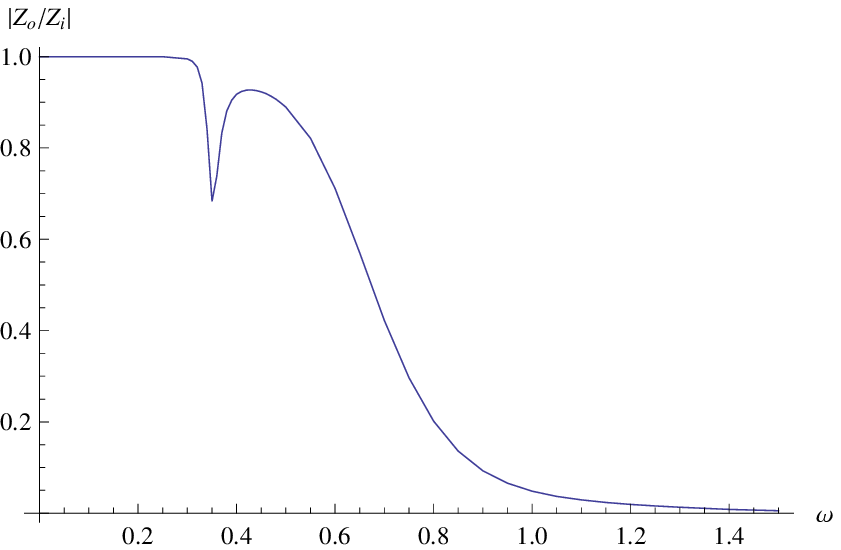}
\caption{The effective potential $V(r)$ (left panel) and the square root of the reflection coefficient $|Z_o/Z_i|$ (right panel) for $D=6$, $\alpha=2$, $l=2$ scalar-type gravitational perturbations.}\label{fig-scalarpot}
\end{figure*}

\begin{widetext}
\section{The wave like equations}\label{sec:wave-like}
\subsection{Gravitational perturbations}

The Einstein-Gauss-Bonnet action is
\begin{equation}
I = \frac{1}{16 \pi G_{D}} \int{d^{D} x \sqrt{-g} R} +  \alpha^\prime
\int d^{D} x \sqrt{-g} (R_{abcd} R^{abcd}- 4 R_{cd}R^{cd} + R^2),
\end{equation}
where $\alpha^\prime$ is the coupling constant; $\alpha^\prime = 1/2 \pi \ell_{s}^{2}$.
The metric has the form,
\begin{equation}\label{metric}
ds^2=f(r)dt^2-\frac{dr^2}{f(r)}-r^2d\Omega_{D-2}^2,
\end{equation}
\begin{equation}
f(r)=1+\frac{r^2\left(1-q(r)\right)}{\alpha(D-3)(D-4)}, \qquad
q(r)=\sqrt{1+\frac{4\alpha(D-3)(D-4)\mu}{(D-2)r^{D-1}}}
\end{equation}
where $d\Omega_{D-2}^2$ is the line element of a unit $(D-2)$-sphere and $\alpha = 16 \pi G_{D} \alpha^\prime$. As we mentioned earlier, we shall measure all quantities in terms of the black hole horizon radius, which we shall denote as $r_0$, so that the black hole mass can be written as
\begin{equation}\label{masseq}
\mu=\frac{(D-2)r_0^{D-3}}{4}\left(2+\frac{\alpha(D-3)(D-4)}{r_0^2}\right).
\end{equation}

The linearized perturbations of (\ref{metric}) can be written in the general form as
\begin{equation} \label{p1}
g_{ab} \to g_{ab} + \delta g_{ab} = g_{ab} + h_{ab}.
\end{equation}
Indices of $h_{ab}$  are raised using the background metric; therefore
$$\delta g^{ab} = - h^{ab}.$$
The first order variation of the Riemann tensors is:
\begin{equation} \label{Rie}
\delta {R_{ab}}^{cd} =  \frac{1}{2} \left\{ {R_{ab}}^{df} {h_f}^c - {R_{ab}}^{cf} {h_f}^d + \left( \nabla_b \nabla^c {h_a}^d - \nabla_a \nabla^c {h_b}^d  \right) + \left( \nabla_a \nabla^d {h_b}^c - \nabla_b \nabla^d {h_a}^c \right) \right\}.
\end{equation}

Using the symmetry of the background metric, one can decompose the perturbation equations into scalars, vectors and tensors according to the rotation group on the $(D-2)$-sphere \cite{Dotti}. Then, the separation of angular variables is possible for this case \cite{Dotti}.

After separation of the angular variables and implying stationarity, $\Psi(t,r)=e^{-\imo\omega t}\Phi(r)$, the dynamics of the gravitational perturbations can be reduced to the wavelike equations \cite{Dotti},
\begin{equation}\label{wave-like}
\left(\frac{d^2}{dr_\star^2} + \omega^2- V_{i}(r)\right)\Phi_{i}(r) = 0\,, \quad dr_\star=\frac{dr}{f(r)},
\end{equation}
with the effective potentials which have a very cumbersome form. After some algebra, we managed to simplify the potentials obtained by Dotti and collaborators in \cite{Dotti} for the tensor, vector, and scalar types of the gravitational perturbation, respectively:

\begin{eqnarray}
\label{tensor-pot}V_T(r)&=&f(r)\frac{\lambda}{r^2}\left(3-\frac{B(r)}{A(r)}\right)+\frac{1}{\sqrt{r^{D-2}A(r)q(r)}}\frac{d^2}{dr_\star^2}\sqrt{r^{D-2}A(r)q(r)},\\
\label{vector-pot}V_V(r)&=&f(r)\frac{(D-2)c}{r^2}A(r)+\sqrt{r^{D-2}A(r)q(r)}\frac{d^2}{dr_\star^2}\frac{1}{\sqrt{r^{D-2}A(r)q(r)}},\\
\label{scalar-pot}V_S(r)&=&\frac{f(r)U(r)}{64 r^2(D-3)^2A(r)^2q(r)^8(4c q(r) + (D-1)R (q(r)^2-1))^2}.\\\nonumber
\end{eqnarray}

Here we used the following dimensionless quantities:
\begin{eqnarray}\nonumber
A(r)&=&\frac{1}{q(r)^2}\left(\frac{1}{2}+\frac{1}{D-3}\right)+\left(\frac{1}{2}-\frac{1}{D-3}\right),\\\nonumber
B(r)&=&A(r)^2\left(1+\frac{1}{D-4}\right)+\left(1-\frac{1}{D-4}\right),\\\nonumber
R&=&\frac{r^2}{\alpha(D-3)(D-4)},
\end{eqnarray}
\begin{eqnarray}\nonumber
U(r)&=&5 (D - 1)^6 R^2 (1 + R) - 3 (D - 1)^5 R ((D - 1) R^2 + 24 c (1 + R)) q(r) + \\\nonumber&& +
 2 (D - 1)^4 (24 c (D - 1) R^2 +
    168 c^2 (1 + R) - (D - 1) R^2 (-3 + 5 R + 7 D (1 + R))) q(r)^2 + \\\nonumber&& +
 2 (D - 1)^4 R (-184 c^2 + (D - 1) (13 + D) R^2 +
    c (-84 + 44 R + 84 D (1 + R))) q(r)^3 + \\\nonumber&& +
    (D - 1)^3 (384 c^3 - 48 c (2 + D (3 D - 5)) R^2 +
    192 c^2 (-11 + D + (-15 + D) R) + \\\nonumber&& + (D - 1) R^2 (-3 (7 + 55 R) +
       D (26 + 106 R + 7 D (1 + R)))) q(r)^4 + \\\nonumber&& +
       (D - 1)^3 R (-64 c^2 (D - 38) + (D - 1) (71 + D (7 D - 90)) R^2 + \\\nonumber&& +
    16 c (303 + 255 R + 13 D^2 (1 + R) - 2 D (73 + 81 R))) q(r)^5 + \\\nonumber&& +
 4 (D - 1)^2 (96 c^3 (-7 + D) -
    8 c (D - 1) (145 - 74 D + 6 D^2) R^2 - \\\nonumber&& -
    8 c^2 (9 - 175 R + D (-58 - 34 R + 11 D (1 + R))) + (D - 1) R^2 (-5 (79 + 23 R) + \\\nonumber&& +
       D (5 (57 + 41 R) + D (-81 - 89 R + 7 D (1 + R))))) q(r)^6 - \\\nonumber&& -
 4 (D - 1)^2 R (8 c^2 (43 + (72 - 13 D) D) + (D - 1) (-63 + D (99 + D (-49 + 5 D))) R^2 + \\\nonumber&& +
    4 c (321 + 465 R + D (121 - 39 R + D (-123 - 107 R + 17 D (1 + R))))) q(r)^7 + \\\nonumber&& +
       (D - 1) (128 c^3 (-9 + D) (D - 5) + 32 c (D - 1) (246 + D (9 + D (-55 + 8 D))) R^2 + \\\nonumber&& +
    64 c^2 (D - 5) (D^2 - 3 + (49 + (D - 4) D) R)  -
    (D - 1) R^2 (1173 + 565 R + \\\nonumber&& + D (-4 (997 + 349 R)  + D (6 (393 + 217 R) + D (-548 - 452 R + 45 D (1 + R)))))) q(r)^8 + \\\nonumber&& +
    (D - 1) R (-64 c^2 (D - 5) (36 + D (-13 + 3 D)) + \\\nonumber&& + (D - 1) (635 + D (-1204 + 3 D (294 + D (-92 + 9 D)))) R^2 - \\\nonumber&& -
    8 c (D - 5) (63 + 31 R + D (127 + 191 R + D (-47 + D + (-79 + D) R)))) q(r)^9 + \\\nonumber&& +
 2 (D - 5) (64 c^3 (D - 5) (D - 3) + 8 c (D - 1) (-27 + D (141 + (-43 + D) D)) R^2 + \\\nonumber&& +
    8 c^2 (D - 5) (-3 + 77 R + D (D - 2 + (D - 18) R)) + (D - 1)^2 R^2 (-33 (R - 7) + \\\nonumber&& +
       D (59 + 43 R + D (-59 - 35 R + 9 D (1 + R))))) q(r)^{10} - \\\nonumber&& -
 2 (D - 5) R (24 c^2 (-11 + D) (D - 5) (D - 3) + (D - 1)^2 (-65 + D (81 + D (7 D - 39))) R^2 + \\\nonumber&& +
    12 c (-7 + D) (D - 5) (D - 3) (D - 1) (1 + R)) q(r)^{11} + \\\nonumber&& +
    (D - 5)^2 (-1 + D) R^2 (16 c (26 + (D - 9) D) + \\\nonumber&& + (D - 1) (77 - 3 R + D (-18 + D + (D - 2) R))) q(r)^{12} + \\\nonumber&& +
    (D - 5)^2 (D - 3)^2 (D - 1)^2 R^3 q(r)^{13},
\end{eqnarray}
$\lambda=(D-2)(c+1)=\ell(\ell + D - 3)$ is the eigenvalue of the angular
part of the Laplacian, $\ell=2,3,4\ldots$.

A wide class of static black holes in the Einstein gravity is known to be stable against gravitational perturbations for any number of space-time dimensions \cite{stabilityD-dim}. This is not true for black holes in the Einstein-Gauss-Bonnet theory which suffer from instability for sufficiently large (in units of the radius of the event horizon) values of the GB coupling $\alpha$ \cite{Dotti,Takahashi:2010gz}. We shall consider here only small enough values of the GB coupling $\alpha$ which are below the threshold of the gravitational instability \cite{Dotti,KonoplyaGBinstability}. In other words, we shall be limited by not very small masses of black holes $M$, which usually are at least one order larger than $M_{*}$.

\subsection{Test scalar field}

We shall also consider the test scalar field in the background (\ref{metric}) which satisfies the Klein-Gordon equation
\begin{equation}\label{Klein-Gordon}
\nabla^a\nabla_a \Phi = 0.
\end{equation}

After separation of the angular variables the equation (\ref{Klein-Gordon}) can be reduced to the wavelike form (\ref{wave-like}) with the following effective potential
\begin{equation}
\label{testscalar-pot}V_{SB}(r)=f(r)\left(\frac{\lambda}{r^2}+\frac{(D-2)(D-4)}{4r^2}f(r)+\frac{(D-2)}{2r}f'(r)\right),
\end{equation}
where $\lambda=\ell(\ell + D - 3)$, $\ell=0,1,2,3,\ldots$.

Let us note, that unlike the Schwarzschild black hole, the Gauss-Bonnet black hole has the effective potential for the scalar field (\ref{testscalar-pot}) which differs from the effective potentials for the tensor-type gravitational perturbations (\ref{tensor-pot}). In the limit $\alpha\rightarrow0$ (\ref{testscalar-pot}) coincides
with (\ref{tensor-pot}).

\subsection{Brane-localized fields}

In addition, we shall consider the Standard Model fields (scalars, fermions and gauge bosons) living on a 4-dimensional brane, which is embedded in the background of the Gauss-Bonnet black hole. The induced metric on the brane is given by a projection of the metric (\ref{metric}) onto the 4-brane \cite{projected}
\begin{equation}\label{projected-metric}
ds^2=f(r)dt^2-\frac{dr^2}{f(r)}-r^2d\Omega_{2}^2,
\end{equation}
where $d\Omega_{2}^2$ is the line element of a unit sphere.

The effective potential for the scalar field (\ref{Klein-Gordon}) in the metric (\ref{projected-metric}) is
\begin{equation}
\label{projscalar-pot}V_{SP}(r)=f(r)\left(\frac{\ell(\ell+1)}{r^2}+\frac{f'(r)}{r}\right).
\end{equation}

The massless gauge field satisfies the equation
\begin{equation}
\nabla^a(\nabla_aA_b-\nabla_bA_a)=0.
\end{equation}

After the separation of the angular variables, one can find the effective potential for the two types of the polarizations
\begin{equation}
\label{projgauge-pot}V_{GP}(r)=f(r)\frac{\ell(\ell+1)}{r^2},
\end{equation}
where $\ell=1,2,3,\ldots$.

For the fermions we do not consider the equation in the standard Schr\"odinger wavelike form because of the boundary conditions: we will calculate the grey-body factors for neutrinos and antineutrinos using the approach of \cite{Casals:2006xp}. That is why we use here the equation for the radial part in the following form

\begin{equation}
\Delta(r)\Psi''(r)+\frac{\Delta'(r)}{2}\Psi'(r)+\left(\frac{2\omega^2r^4-\imo\omega r^2\Delta'(r)}{2\Delta(r)}+2\imo\omega r-\kappa^2\right)\Psi(r)=0,
\end{equation}
where $\Delta(r)=r^2f(r)$ and $\kappa=1,2,3\ldots$.
\end{widetext}

\section{Calculations of the energy-emission rate}\label{sec:methods}
In order to calculate the intensity of the Hawking radiation, one should first calculate the gray-body factors, that is, to solve the problem of classical scattering around black holes with pure in-going boundary conditions at the event horizon. The latter is reduced to the finding of the S-matrix, or, simply, of the reflection or the transmission coefficients.

\subsection{Reflection coefficients}

At the event horizon we impose the boundary condition that corresponds to the purely ingoing wave
\begin{equation}\label{horizon-BC}
\Phi(r)\propto e^{-\imo\omega r_\star}\propto
\left(r-r_0\right)^{-\imo\omega/f'(r_0)}.
\end{equation}
At the spatial infinity ($r\rightarrow\infty$) the two linearly independent solutions of the wavelike equation
(\ref{wave-like}) are
$$\Phi(r)\simeq Z_i \exp(-\imo\omega r_\star)+Z_o \exp(\imo\omega r_\star),$$
where $Z_i$ and $Z_o$ are integration constants which correspond
to the ingoing and outgoing waves respectively.
Introducing the new function
$$P(r)=(1 - (r_0/r))^{\imo\omega/f'(r_0)}\Phi(r),$$
and choosing the integration constant as $P(r_0)=1,$ we expand the equation (\ref{wave-like}) near the event horizon and find $P'(r_0)$, which completely determines initial conditions for the numerical integration. Then, we integrate numerically the equation (\ref{wave-like}) from the event horizon $r_0$ until some distant point $R\gg r_0$ and find a fit for the numerical solution far from the black hole in the form
\begin{equation}\label{fit}
P(r)=Z_i P_i(r)+Z_o P_o(r),
\end{equation}
where the asymptotical expansions for the corresponding functions are found by expanding
(\ref{wave-like}) for large $r$ as
\begin{eqnarray}
P_i(r)&=&e^{-\imo\omega r}\left(1+P_i^{(1)}r^{-1} + P_i^{(2)}r^{-2}+\ldots\right),\nonumber\\
P_o(r)&=&e^{\imo\omega r}\left(1+P_o^{(1)}r^{-1} + P_o^{(2)}r^{-2}+\ldots\right).\nonumber
\end{eqnarray}

The fitting procedure allows us to find the coefficients $Z_i$ and $Z_o$. In order to check the accuracy of the calculated coefficients one should increase the internal precision of the numerical integration procedure, the value of $R$ and the number of terms in the series expansion for $P_i(r)$ and $P_o(r)$, making sure that the values of $Z_i$ and $Z_o$ do not change within the desired precision.

\subsection{WKB approach}\label{sec:WKB}
Another way to check our numerical calculations is to compare the fraction $Z_o/Z_i$ with the result provided by the 6-th order WKB method, which gives quite an accurate answer for large and moderate values of $\ell$ \cite{WKB}. The WKB approach was initially used for finding quasinormal modes for which it usually provides quite a good accuracy at the 6th order \cite{WKBuse}

The reflection coefficient, given by the WKB formula, is
\begin{equation}\label{moderate-omega-wkb}
|Z_o/Z_i|^2 = (1 + e^{- 2 i \pi K})^{-1},
\end{equation}
where
\begin{equation}
K = i \frac{(\omega^2 - V_{0})}{\sqrt{-2 V_{0}^{\prime \prime}}} + \sum_{i=2}^{i=6} \Lambda_{i}.
\end{equation}
Here $V_0$ is the maximum of the effective potential, $V_{0}^{\prime \prime}$ is the second derivative of the effective potential in its maximum with respect to the tortoise coordinate, and $\Lambda_i$  are higher order WKB corrections which depend on up to the $2i$-th order derivatives of the effective potential at its maximum.

From Fig. \ref{fig-WKBcheck} we see that the difference between the results found by these two methods is a fraction of a percent. Unfortunately, for some values of the parameters the effective potential does not have the form of a peak (see Fig. \ref{fig-scalarpot}). In these cases the WKB formula cannot be used in the present form. That is why, despite in most cases the WKB formula provides a good approximation, we use a numerical integration procedure which works for any form of the effective potential.

\begin{figure}
\includegraphics[width=\linewidth,clip]{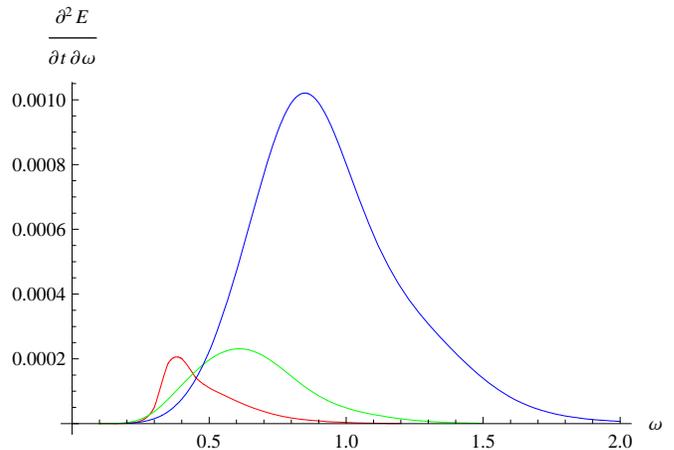}
\caption{The energy-emission rates for the Gauss-Bonnet black holes $D=6$, $\alpha=1/5$ (blue, top line), $\alpha=1/2$ (green, middle line), and $\alpha=1$ (red, bottom line).}\label{fig-D6}
\end{figure}

\begin{figure*}
\includegraphics[width=.5\textwidth,clip]{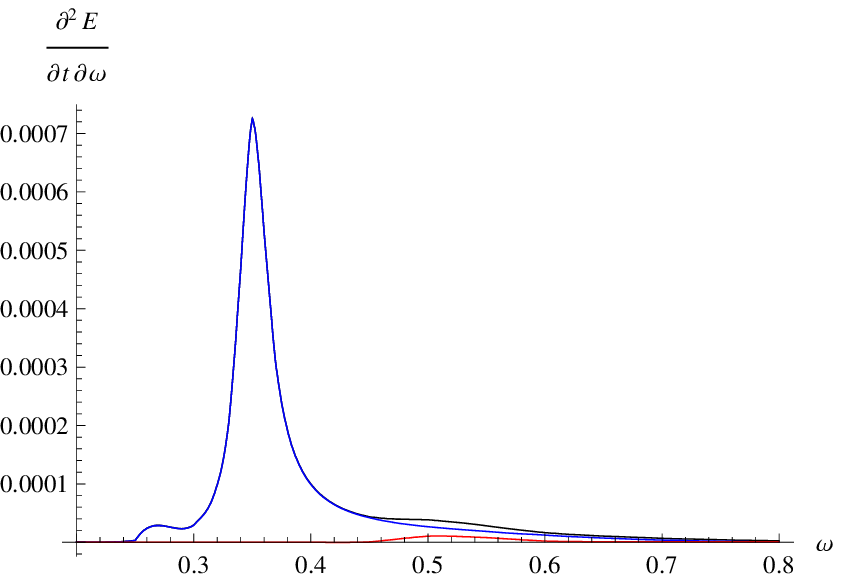}\includegraphics[width=.5\textwidth,clip]{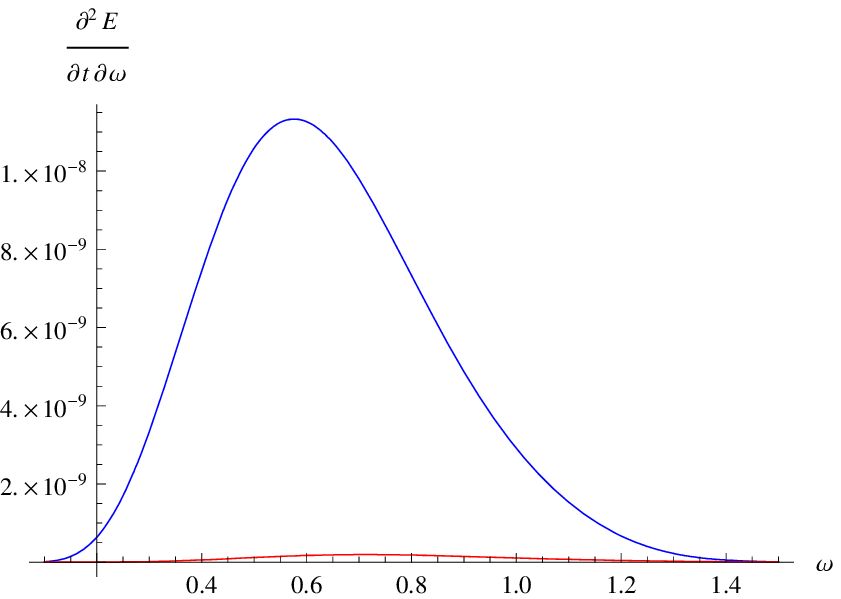}
\caption{The energy-emission rate (left panel) and the contribution of the graviton emission of a tensor-type (right panel) for the Gauss-Bonnet black hole $D=6$, $\alpha=2$. Blue (top) and red (bottom) lines correspond respectively to the contribution of $l=2$ and $l=3$ multipoles of a scalar-type (left panel) and of a tensor-type (right panel).}\label{fig-D6a2}
\end{figure*}

\begin{figure*}
\includegraphics[width=.33\textwidth,clip]{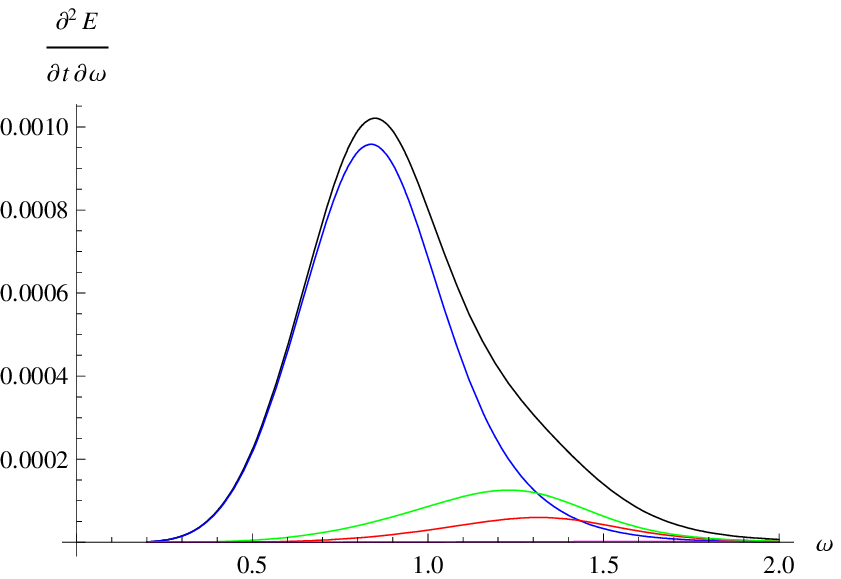}\includegraphics[width=.33\textwidth,clip]{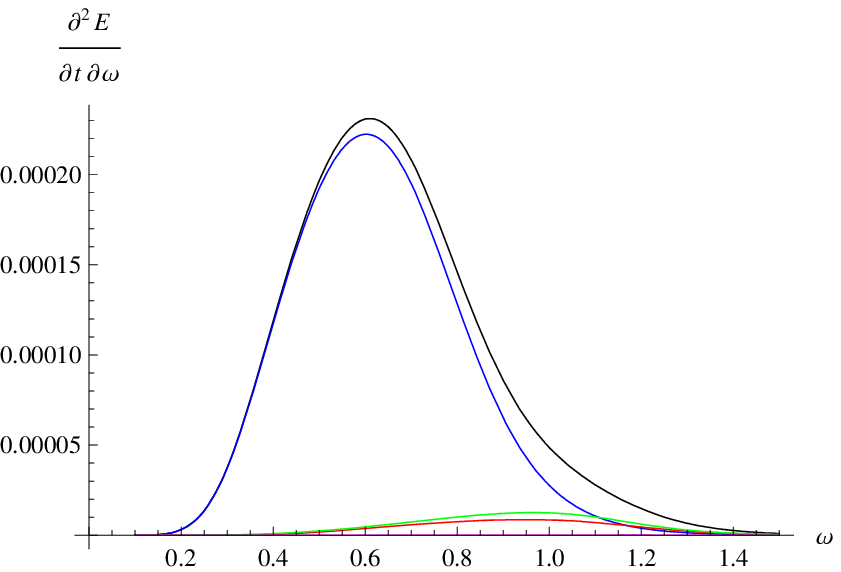}\includegraphics[width=.33\textwidth,clip]{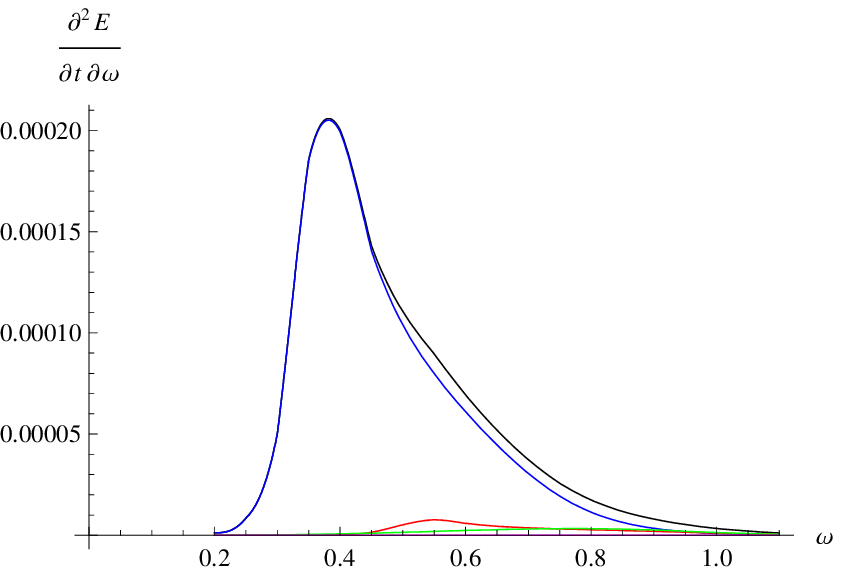}
\caption{The energy-emission rates (black line) for the Gauss-Bonnet black holes $D=6$, $\alpha=1/5,1/2,1$ (from left to right) together with contributions of different types of gravitons (colored lines): $l=2$ scalar type (blue, top), $l=2$ vector type (green), $l=3$ scalar type (red), and $l=2$ tensor type (magenta, bottom).
}\label{fig-D6contrib}
\end{figure*}

\begin{figure*}
\includegraphics[width=.5\textwidth,clip]{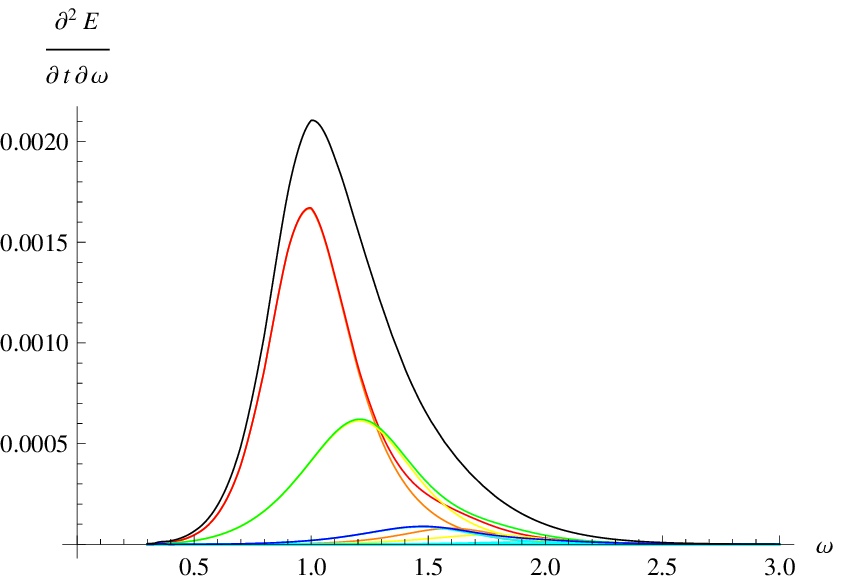}\includegraphics[width=.5\textwidth,clip]{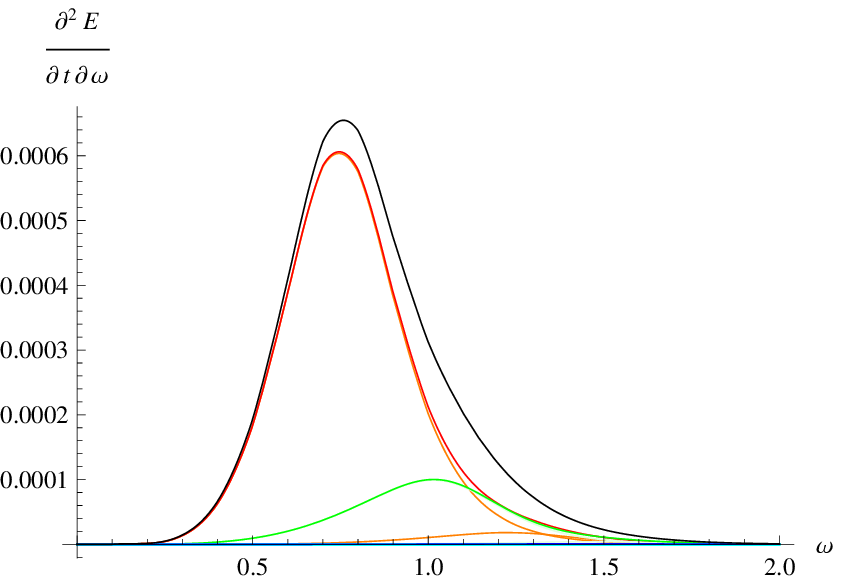}
\caption{The energy-emission rates for $D=5$ Schwarzschild (left panel) and Gauss-Bonnet black holes ($\alpha=1/5$, right panel). The black line is the total energy-emission rate. Scalar-type, vector-type, and tensor-type gravitons' contributions are red, green, and blue. The contributions of the corresponding multipole numbers are orange, yellow, and cyan. The largest gravitons' contribution is of a scalar-type, the smallest is of a tensor-type.}\label{fig-D5}
\end{figure*}

\begin{figure*}
\includegraphics[width=.5\textwidth,clip]{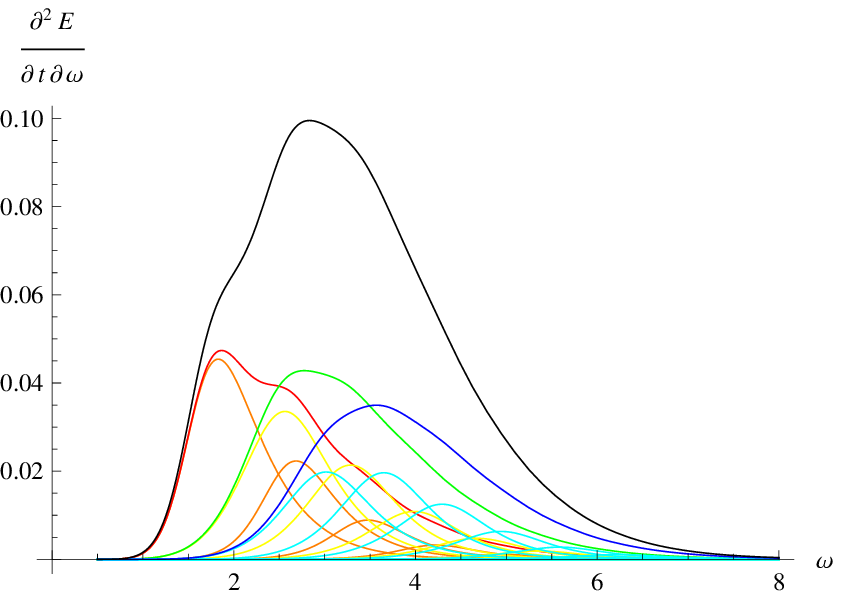}\includegraphics[width=.5\textwidth,clip]{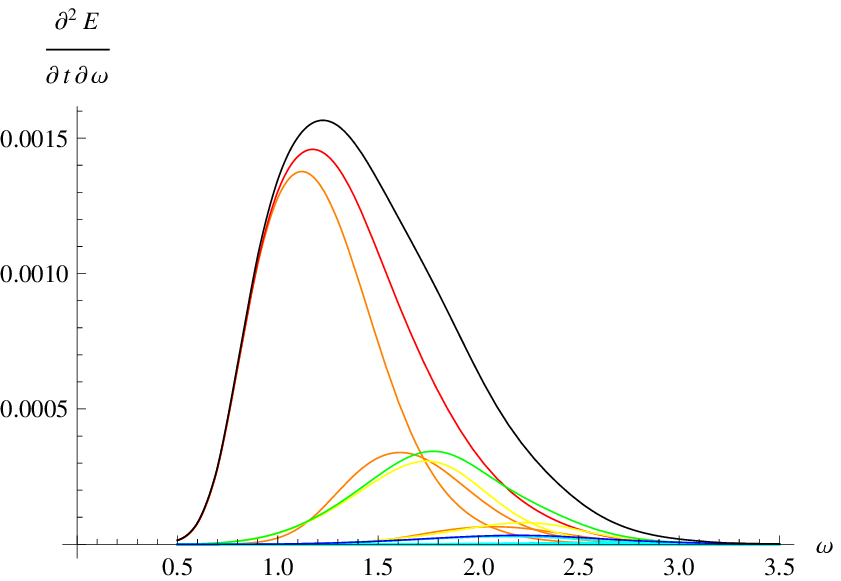}
\caption{The energy-emission rates for $D=8$ Schwarzschild (left panel) and Gauss-Bonnet black holes ($\alpha=1/5$, right panel). The black line is the total energy-emission rate. Scalar-type, vector-type, and tensor-type gravitons' contributions are red, green, and blue. The contributions of the corresponding multipole numbers are orange, yellow, and cyan. The largest gravitons' contribution is of a scalar-type, the smallest is of a tensor-type.}\label{fig-D8}
\end{figure*}

\begin{figure*}
\includegraphics[width=.5\textwidth,clip]{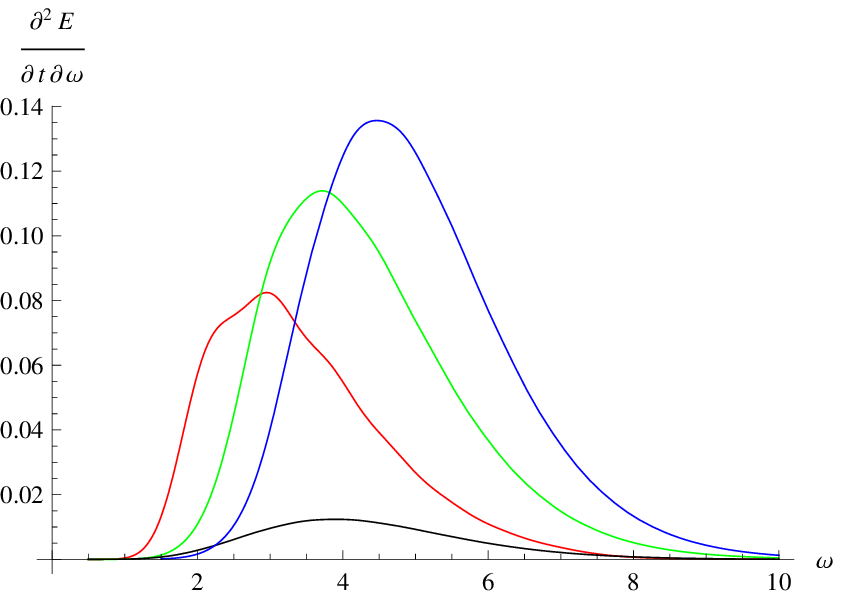}\includegraphics[width=.5\textwidth,clip]{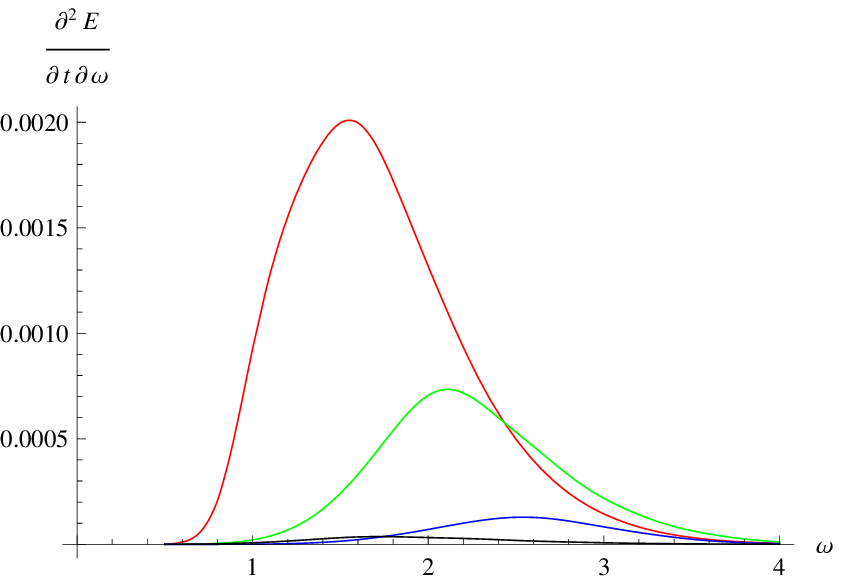}
\caption{The energy-emission rates for $D=9$ Schwarzschild (left panel) and Gauss-Bonnet black holes ($\alpha=1/5$, right panel). Scalar-type, vector-type, and tensor-type gravitons' emission rates are red, green, and blue. For the Schwarzschild black hole the largest gravitons' contribution is of tensor-type and the smallest is of scalar-type, while for $\alpha=1/5$ Gauss-Bonnet black hole the largest gravitons' contribution is of a scalar-type and the smallest is of a tensor-type. The black (bottom) line is the energy-emission rate due to the scalar field.}\label{fig-D9}
\end{figure*}

\begin{figure*}
\includegraphics[width=.5\textwidth,clip]{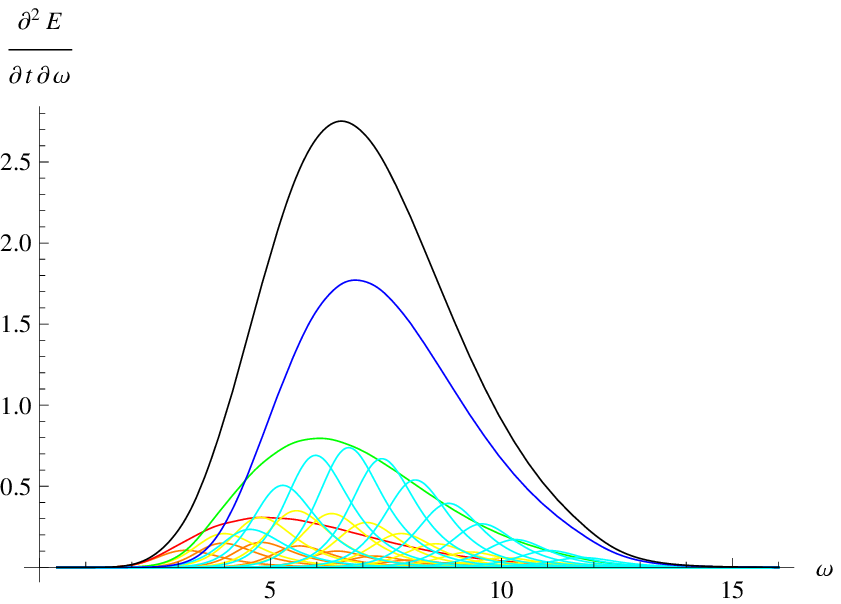}\includegraphics[width=.5\textwidth,clip]{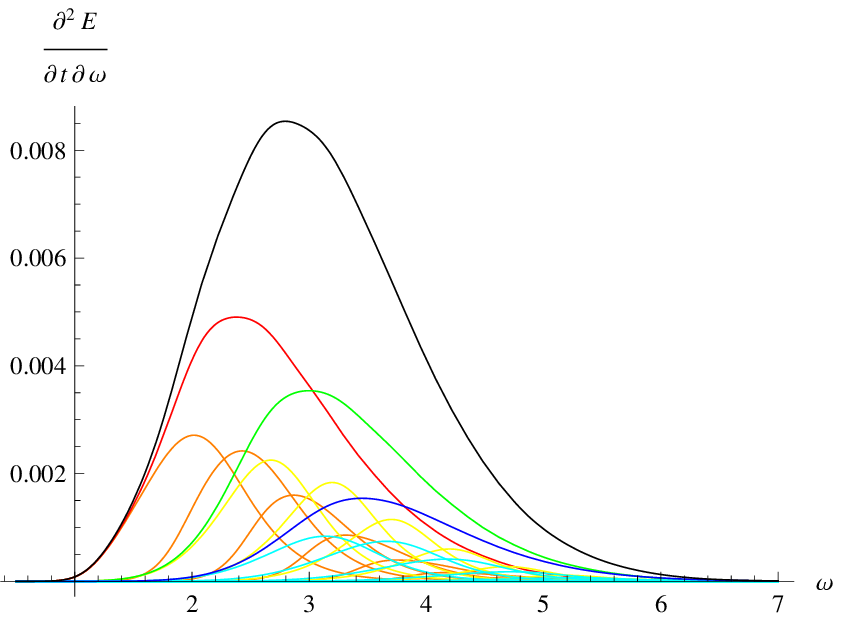}
\caption{The energy-emission rates for $D=11$ Schwarzschild (left panel) and Gauss-Bonnet black holes ($\alpha=1/5$, right panel). The black line is the total energy-emission rate. Scalar-type, vector-type, and tensor-type gravitons' contributions are red, green, and blue. The contributions of the corresponding multipole numbers are orange, yellow, and cyan. For the Schwarzschild black hole the largest gravitons' contribution is of a tensor-type and the smallest is of a scalar-type, while for $\alpha=1/5$ Gauss-Bonnet black hole the largest gravitons' contribution is of a scalar-type and the smallest is of a tensor-type.}\label{fig-D11}
\end{figure*}

\begin{figure*}
\includegraphics[width=.5\textwidth,clip]{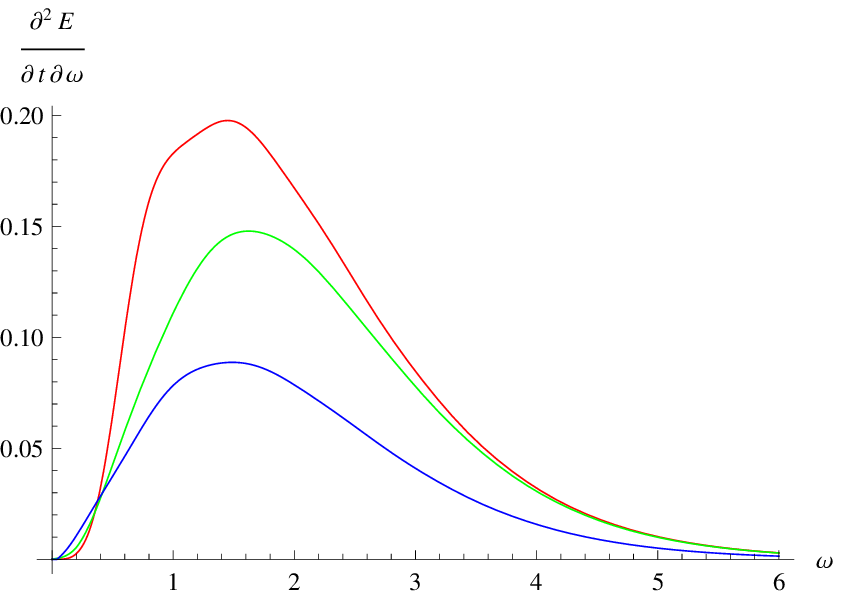}\includegraphics[width=.5\textwidth,clip]{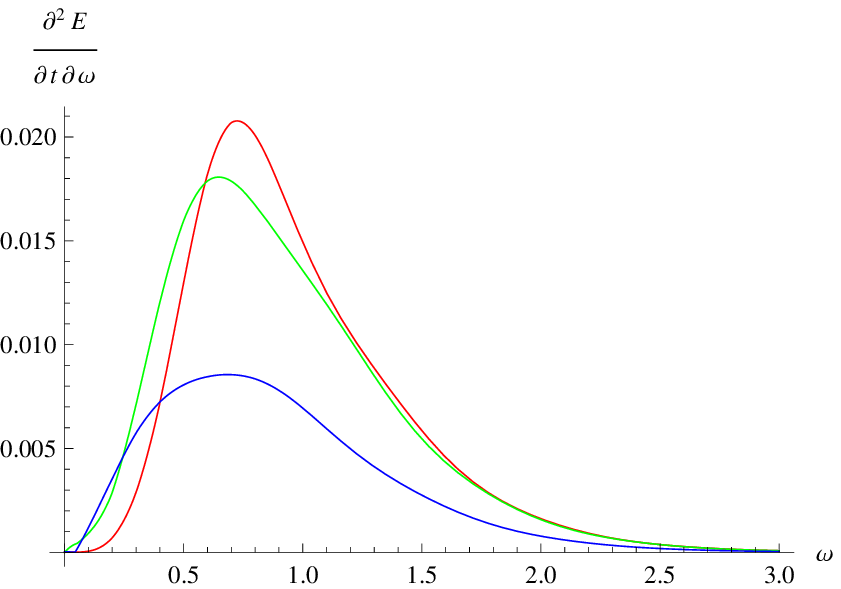}
\caption{The energy-emission rates for $D=10$ Schwarzschild (left panel) and Gauss-Bonnet black holes ($\alpha=1/5$, right panel) due to brane-localized fields: gauge (red, top), Dirac (green, middle), and scalar (blue, bottom).}\label{fig-D10}
\end{figure*}

\subsection{Graviton emission}
When the coefficients $Z_i$ and $Z_o$ are obtained, one can find the absorption probability
\begin{equation}\label{absorbtion}
|{\cal A_\ell }|^2=1-|Z_o/Z_i|^2
\end{equation}
and, then, the energy-emission rate
\begin{equation}\label{energy-emission}
{{dE} \over {dt}} =
\sum_\ell {N_\ell \left| {\cal A_\ell } \right|^2 {\omega \over
{\exp (\omega /T_H ) - 1}}{{d\omega } \over {2\pi }}},
\end{equation}
where the multiplicity factors $N_\ell$ are \cite{Kanti:2009sn}
\begin{eqnarray}
\nonumber N_\ell  ^{(T)} &=&
{{(D\!-\!1)(\ell\!+\!D\!-\!2)(\ell\!-\!1)(2\ell\!+\!D\!-\!3)(\ell\!+\!D\!-\!5)!}\over{2(\ell\!+\!1)!(D\!-\!3)(D\!-\!5)!}}\,,\\
\nonumber N_\ell^{(V)} &=& {{(\ell+D-3)\ell (2\ell+D-3)(\ell+D-5)!} \over {(\ell+1)!(D-4)!}}\,,\\
\nonumber N_\ell^{(S)} &=& {{(2\ell+D-3)(\ell+D-4)!} \over {\ell!(D-3)!}}\,,
\end{eqnarray}
for the gravitational perturbations of tensor, vector and scalar
types. The Hawking temperature is 
$$T_H=\frac{f'(r_0)}{4\pi}=\frac{(D-3) (2 r_0^2+ \alpha(D-4)(D-5))}{8\pi r_0 (r_0^2 + \alpha(D-3)(D-4))}.$$

\subsection{Emission of the scalar field and the Standard Model fields}
The scalar field living in the bulk has the same number of the degrees of freedom as the gravitational perturbations of scalar type
\begin{equation}
\nonumber N_\ell^{(SB)} = {{(2\ell+D-3)(\ell+D-4)!} \over {\ell!(D-3)!}}\,.
\end{equation}
For the scalar field localized on the brane the multiplicity factor is given
\begin{equation}
\nonumber N_\ell^{(SP)} = {{(2\ell+1)\ell!} \over {\ell!1!}}=2\ell+1\,.
\end{equation}

For the gauge field the multiplicity factor is the same for each polarization
\begin{equation}
\nonumber N_\ell^{(GP)} = 2\ell+1\,.
\end{equation}

For the neutrinos and antineutrinos the multiplicity factor is
\begin{equation}
\nonumber N_\kappa^{(FP)} = 2\kappa\,
\end{equation}

In order to calculate the absorption probability for the fermions we use the approach of \cite{Casals:2006xp}
$$|{\cal A_\kappa }|^2=1-\frac{4\omega^2}{\kappa^2}\left|\frac{Z_o}{Z_i}\right|^2.$$
The energy-emission rate for the neutrinos and antineutrinos is the same and given by
$${{dE} \over {dt}} =
\sum_{\kappa=1}^\infty {N_\kappa^{(FP)} \left| {\cal A_\kappa } \right|^2 {\omega \over
{\exp (\omega /T_H ) + 1}}{{d\omega } \over {2\pi }}}.
$$

Note that we take into account contributions of neutrinos and antineutrinos and both polarizations of the gauge bosons. That is why our result is two times larger than in \cite{Cardoso:2005vb}.

\begin{table*}
\caption{Energy-emission rate for Schwarzschild (first line) and Schwarzschild-Gauss-Bonnet ($\alpha/r_0^2=1/5$) (second line) black holes.}\label{thetable}
\begin{tabular}{|r||l|l|l|l||l|l|l|l|}
\hline
D&\multicolumn{4}{c||}{gravitational}&\multicolumn{2}{c|}{scalar field}&\multicolumn{2}{c|}{projected SM fields}\\
\cline{2-9}
&total&scalar&vector&tensor&bulk&projected&$s=1/2$&$s=1$\\
\hline
$5$&$0.00133$&$0.00085$&$0.00041$&$0.00006$&$0.00105$&$0.00266$&$0.00464$&$0.00365$\\
   &$0.000322$&$0.000267$&$0.000055$&$0.000001$&$0.000230$&$0.000772$&$0.001259$&$0.000733$\\
\hline
$6$&$0.01380$&$0.00796$&$0.00437$&$0.00147$&$0.00267$&$0.01073$&$0.01955$&$0.01944$\\
   &$0.000618$&$0.000535$&$0.000082$&$0.000002$&$0.000100$&$0.001141$&$0.002051$&$0.001340$\\
\hline
$7$&$0.06942$&$0.03229$&$0.02366$&$0.01347$&$0.00648$&$0.02972$&$0.05310$&$0.05935$\\
   &$0.000994$&$0.000833$&$0.000152$&$0.000008$&$0.000057$&$0.001832$&$0.003440$&$0.002528$\\
\hline
$8$&$0.27143$&$0.09152$&$0.09648$&$0.08343$&$0.01614$&$0.06613$&$0.11527$&$0.13742$\\
   &$0.001737$&$0.001366$&$0.000332$&$0.000038$&$0.000047$&$0.003197$&$0.006157$&$0.005043$\\
\hline
$9$&$0.9975$&$0.2294$&$0.3440$&$0.4241$&$0.04216$&$0.12769$&$0.21768$&$0.26977$\\
   &$0.003490$&$0.002430$&$0.000897$&$0.000162$&$0.000050$&$0.005631$&$0.010781$&$0.009809$\\
\hline
$10$&$3.6665$&$0.5592$&$1.1639$&$1.9434$&$0.11595$&$0.22321$&$0.37362$&$0.47460$\\
    &$0.007727$&$0.004550$&$0.002425$&$0.000752$&$0.000064$&$0.009633$&$0.018190$&$0.017976$\\
\hline
$11$&$13.7745$&$1.4696$&$3.8936$&$8.4113$&$0.34326$&$0.36257$&$0.59793$&$0.77210$\\
    &$0.018552$&$0.008939$&$0.006678$&$0.002935$&$0.000094$&$0.015795$&$0.029331$&$0.030885$\\
\hline
\end{tabular}
\end{table*}

\section{Results}\label{sec:results}

In Fig. \ref{fig-D6} one can see that the energy-emission rate per unit frequency $\omega$ for gravitons strongly decreases as $\alpha/r_{0}^2$ increases, leading to suppression of about five times for $\alpha/r_{0}^2 =1$. Examples of contributions of various multipoles and of various types of gravitons (scalar, vector and tensor) are given in Figs. \ref{fig-D6a2}, \ref{fig-D6contrib}, \ref{fig-D5}, \ref{fig-D8}, \ref{fig-D9}, \ref{fig-D11}. These are given for $D=6$ as an example, while from the table \ref{thetable}, we can see the relative contributions for all types of particles for various numbers of space-time dimensions. In Fig. \ref{fig-D10} one can see contributions of brane-localized Standard model fields into the energy-emission rate: for $D\geq6$ the larger spin of a field, the larger is the corresponding energy-emission rate per unit frequency around its maximum.

An important observation for our future discussion is about the role of the tensor type of gravitons in the evaporation process. For vanishing GB coupling the tensor type of gravitons corresponds to the highest energy-emission rate, which leads to the dominance of tensorial gravitons among all particles in the Hawking radiation at large $D$ (see Figs. \ref{fig-D9}, \ref{fig-D11}). For non zero $\alpha$ this is not true anymore: the energy-emission rates of tensorial gravitons become strongly suppressed and the Standard Model particles dominate in the evaporation at high $D$.

\begin{figure}
\centerline{\includegraphics[width=\linewidth,clip]{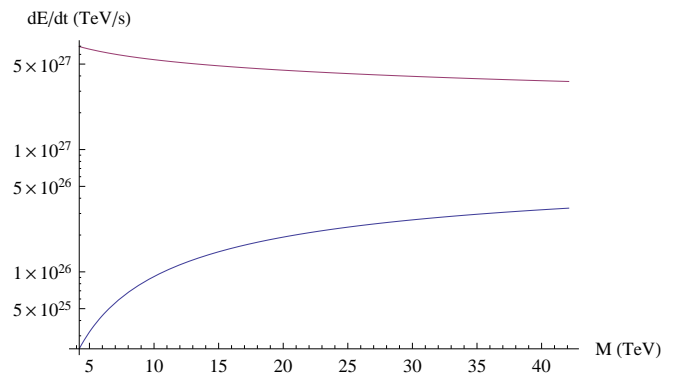}}
\caption{The energy-emission rate as a function of mass $M$ for $D=10$ Schwarzschild (top line) and $\alpha=0.1TeV^{-2}$ GB (bottom line) black holes, $M_*=1TeV$. According to the recent experiments $M_*\geq0.6TeV$ (for $D=10$) \cite{Kanti:2008eq}. For larger $M_*$ the plots are qualitatively the same.}\label{fig-eermass}
\end{figure}


Finally, let us justify here in more detail why we chose the units of the fixed event horizon and not the units of the fixed black hole mass.

First, we would like to mention that from the values in units of the black hole radius, using a simple formula which we shall show in the next section, we are always able to recalculate values of the energy-emission rate in units of the black hole mass or Planck mass. The inverse formula is not so simple because the mass appears in the left-hand side as well. Second, in order to calculate the energy-emission rate in units of the mass, we must fix in some way the real Planck mass, which we do not yet know. Third, the purpose of the table \ref{thetable} is to show that, independently on the new Planck mass, the energy-emission rate decreases some orders. This is a general result, which we would lose if we fix the Planck mass in some way. In order to show this effect we present a figure in the units of TeVs to illustrate how the emission rate depends on the back hole mass (Fig. \ref{fig-eermass}). 

\begin{figure}
\includegraphics[width=\linewidth,clip]{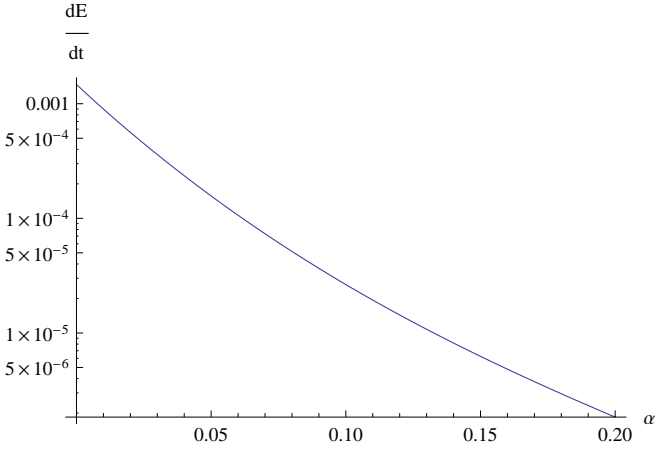}
\caption{Energy-emission rate of the tensor-type gravitons as a function of $\alpha$ for $D=6$.}\label{fig-tensor}
\end{figure}

\begin{figure}
\centerline{$V(r)=$\includegraphics[width=.3\linewidth,clip]{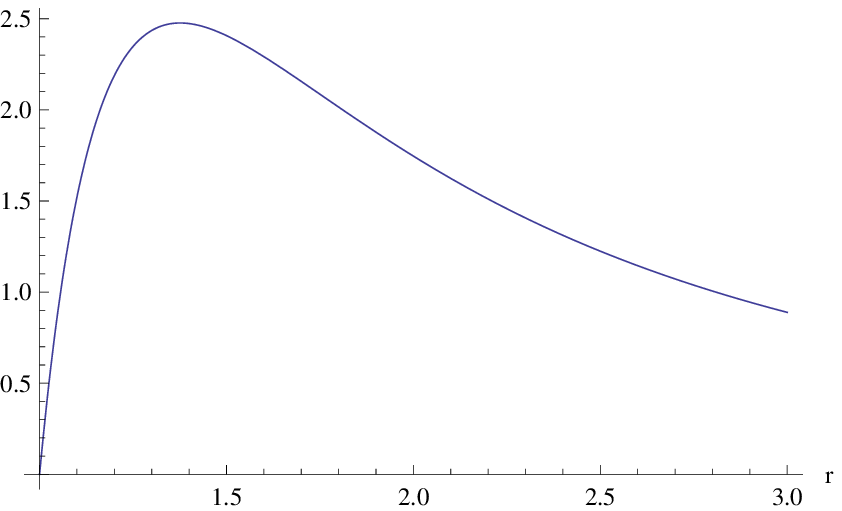}$+\alpha\times$\includegraphics[width=.3\linewidth,clip]{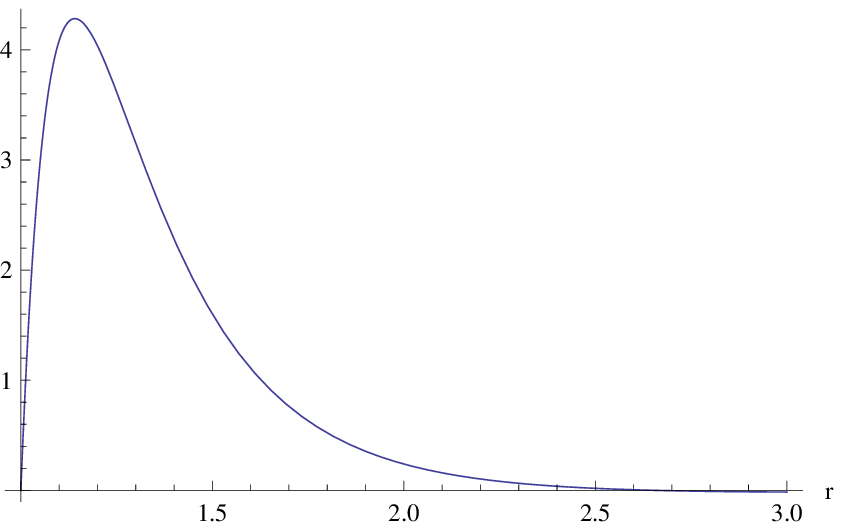}$+{\cal O}(\alpha^2).$}
\caption{Expansion of the tensor-type perturbation potential $D=5$, $\ell=2$.}\label{fig-potential}
\end{figure}

\begin{figure}
\includegraphics[width=\linewidth,clip]{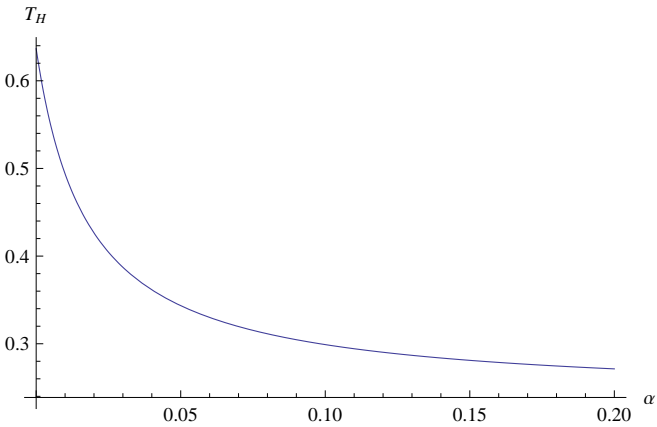}
\caption{Black hole temperature as a function of $\alpha$ for $D=11$.}\label{fig-temperature}
\end{figure}

\begin{figure}
\includegraphics[width=\linewidth,clip]{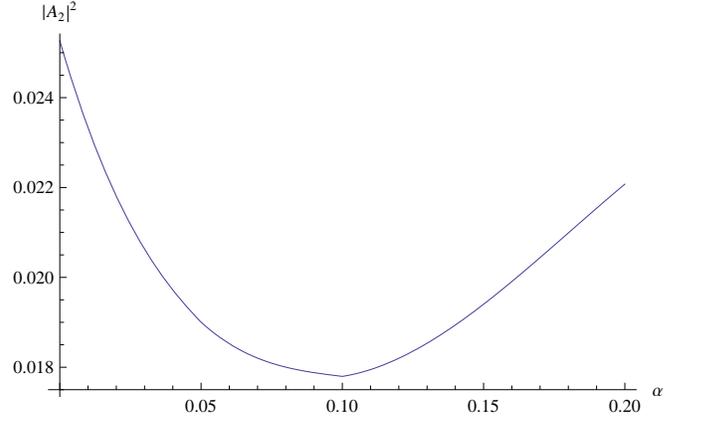}
\caption{Grey-body factor of the tensor-type gravitational perturbations as a function of $\alpha$ ($D=6$, $\ell=2$, $\omega r_0=1.5$).}\label{fig-greybody}
\end{figure}

\section{Discussions}\label{sec:discussions}

From Fig. \ref{fig-tensor} and Table \ref{thetable} one can see that the energy-emission rate decreases quickly (in fact, almost exponentially) with $\alpha$ . In order to understand why this happens for relatively small corrections to the black hole geometry let us use the WKB approach and consider
the cases of small and large $\omega$ of separately.

The numerical data, which we obtained here (Table \ref{thetable}), do not use the small $\omega$ expansion of \cite{Cardoso:2005vb} and are, thereby, more accurate.

When $\omega^2<V_{max}$, the energy-emission rate decreases due to the suppression of the grey-body factor for tensor gravitons. In order to show this we use the WKB approximation. For $\omega^2\ll V_{max}$ the WKB formula reads
\begin{equation}\label{WKBformula}
|{\cal A}|^2\approx \exp\left(-2\int dr_*\sqrt{V(r_*)-\omega^2}\right),
\end{equation}
where integration is performed between the two turning points $V(r_*)=\omega^2$. On Fig. \ref{fig-potential} we see that the leading order of $\alpha$ increases the height of the potential barrier, as well as the distance between the turning points. Then, from (\ref{WKBformula}) it is clear that that the grey-body factor decreases exponentially as the effective potential for tensor gravitons grows.

In the eikonal approximation for $\omega^2\lesssim V_{max}$ one has
$$|{\cal A}|^2\approx \frac{e^{-2\pi(V_{max}-\omega^2)/\sqrt{-2V_{max}''}}}{1+e^{-2\pi(V_{max}-\omega^2)/\sqrt{-2V_{max}''}}}.$$
As the height of the potential increases with growing $\alpha$, the dominant contribution comes from the numerator and the grey-body factor for tensor gravitons is
$$|{\cal A}|^2\propto e^{-C(\omega)\alpha},\quad C(\omega)>0.$$
For large values of $\omega$ it is evident that $|{\cal A}|^2\approx1$. Thus, the dominant contribution to the energy-emission rate at large $\omega$ comes from the temperature, which decreases as (see Fig. \ref{fig-temperature})
$$T_H=T_H(0)\left(1 - \frac{(D-1)(D-4)\alpha}{2r_0^2}\right).$$
Therefore the energy-emission rate decreases exponentially with $\alpha$
$${{\partial^2E} \over {\partial t\partial\omega}} \approx
\sum {N_\ell
{{\omega} \over {2\pi }}}e^{-\omega /T_H }{{\partial^2E} \over {\partial t\partial\omega}}\Biggr|_{\alpha=0}\PRDonly{\!\!\!\!\!\!\!\!\!\!}\times e^ {-\omega(D-1)(D-4)\alpha/(2r_0^2T_H)}.$$

Let us compare the Hawking radiation of Schwarzschild and GB black holes produced due to particle collisions of the same energy. The relation between the black hole mass $M$ and its radius $r_0$ is given by
\begin{equation}
1\!+\!\frac{(D-3)(D-4)\alpha}{2r_0^2}\!=\!\frac{8(M/M_*)\Gamma(\frac{D-1}{2})}{(r_0 M_*)^{D-3} (D-2)\pi^{(D-3)/2}},
\end{equation}
where $M_*$ is the true fundamental scale of gravity. In units of the black hole horizon the energy-emission rate $\sim r_0^{-2}$. If we measure the black hole radius in units of $TeV^{-1}$, the energy-emission rate is measured in units of $TeV^2$. In order to convert the energy-emission rate in the units of $TeV/s$, we divide it by the Plank constant $\hbar=6.58\cdot 10^{-28}TeV\cdot s$. For $\alpha = 0.1 M_{*}^{-2}$ and $D=10$, the total energy-emission rate of GB black holes can be as much as $10^3$ times smaller (see Fig. \ref{fig-eermass}) than that of the Schwarzschild one. For other values of $D$ the increase in the lifetime of black holes can be easily calculated in the same way.

Thus, we have shown that even small GB corrections to the $D$-dimensional Schwarzschild geometry lead to the great increasing of the lifetime of black holes, up to quite a few orders. This is certainly not enough for accreting of matter and thus is not dangerous for experiments at the LHC, yet, it will produce a potentially observable effect.

A natural question is, what is the fraction of the contribution of the grey-body factor into suppression of emission of the tensorial gravitons, and consequently into overall lifetime of black holes. An exact answer is, however, not as easy as it drastically depends on a number of conditions: the mass of the black hole $M$, the value of the GB coupling $\alpha$, the number of space-time dimensions $D$ and is strongly connected with the dominance of a particular type of particles for fixed $\alpha$, $M$, $D$.

A brief hint could be given by the plot of the grey-body factor as a function of $\alpha$, which is shown in Fig. \ref{fig-greybody}. For example, the grey-body factor for $D=6$ can be suppressed at the rate of about 35\% for sufficiently small $\alpha/r_0^2 \approx 0.1$, which is one tenth of its threshold values of instability \cite{Dotti}. Going beyond very small values of $\alpha$ would require inclusion of corrections of higher orders of curvature in the action. Thus, the first factor of the two which was mentioned in the abstract of this article, i.e. the quick cooling of a black hole when $\alpha$ grows, is definitely a dominant factor of suppression of the Hawking radiation.

Let us note that the suppression of the energy-emission rate at high $\omega$ due to the cooling of a black hole was observed in \cite{Rizzo:2006uz} for the Standard Model particles. Though the largest suppression of the graviton emission reported in \cite{Rizzo:2006uz} is at $M \lesssim M_{*}$. In this regime however one cannot trust the solution (\ref{metric}), which is gravitationally unstable \cite{Dotti}. The other factor, the decreasing of the grey-body factors of the tensorial gravitons, seemingly increased the suppression of the graviton emission at small $\alpha$.

\section{Conclusions}\label{sec:conclusions}
We have shown that the widely accepted approximation of higher dimensional black holes by their classical Schwarzschild-Tangherlini model is not good, when one considers the Hawking radiation around a black hole. Intensive Hawking emission of gravitons, as well as of other particles, is suppressed by many orders, when one takes into consideration small quantum Gauss-Bonnet corrections. Consequently, the lifetime of quantum corrected black holes is many orders larger than it is expected according to the current literature \cite{Hawking-raznoe}. This makes further investigations of Hawking radiation of higher curvature corrected black holes appealing.

\section*{Acknowledgments}
R. K. was supported by \emph{the Alexander von Humboldt Foundation}, Germany.
A. Z. was supported by \emph{Funda\c{c}\~ao de Amparo \`a Pesquisa do Estado de S\~ao Paulo (FAPESP)}, Brazil.

\end{document}